\documentclass{aa} 
\usepackage{float} 
\usepackage{graphicx}  
\usepackage{txfonts}  
\usepackage{natbib}
\bibpunct{(}{)}{;}{a}{}{,}
\usepackage{color}
\usepackage[normalem]{ulem}

\begin{document}  

% \bibpunct{(}{)}{;}{a}{}{,} % to follow the A&A style 
 
\title{Asteroseismology of the exoplanet-host F-type star 94 Ceti : impact of atomic diffusion on the stellar parameters }
  
  \author{ M. Deal\inst{1,3}, M.E. Escobar\inst{2,4}, S. Vauclair\inst{2,3}, G. Vauclair\inst{2,3}, A. Hui-Bon-Hoa\inst{2,3}, \and O. Richard\inst{1}}
\institute{Laboratoire Univers et Particules de Montpellier (LUPM), UMR 5299, Universit\'e de Montpellier, CNRS, Place Eug\`ene Bataillon 34095 Montpellier Cedex 5 FRANCE\     
           \and
           Universit\'e de Toulouse, UPS-OMP, IRAP, France
           \and
           CNRS, IRAP, 14 avenue Edouard Belin, 31400 Toulouse, France\
		\and
		Programa Explora, CONICYT, Ministerio de Educación, Moneda 1375, Santiago, Chile\\
            \email{morgan.deal@umontpellier.fr} 
           }
           
\date{\today}

\abstract
% Context
{A precision of order one percent is needed on the parameters of exoplanet-hosts stars in order to correctly characterize the planets themselves. This will be achieved by asteroseismology. It is important in this context to test the influence on the derived parameters of introducing atomic diffusion with radiative accelerations in the models. In this paper, we begin this study with the case of the star 94 Ceti A.}
% Aims
{We performed a complete asteroseismic analysis of the exoplanet-host F-type star 94 Ceti A, from the first radial-velocity observations with HARPS up to the final computed best models. This star is hot enough to suffer from important effects of atomic diffusion, including radiative accelerations. We tested the influence of such effects on the computed frequencies and on the determined stellar parameters. We also tested the effect of including a complete atmosphere in the stellar models.}
% Method
{The radial velocity observations were done with HARPS in 2007. The low degree modes were derived and identified using classical methods and compared with the results obtained from stellar models computed with the Toulouse Geneva Evolution Code (TGEC).}
% Results
{We obtained precise parameters for the star 94 Ceti A. We showed that including atomic diffusion with radiative accelerations can modify the age by a few percents, whereas adding a complete atmosphere does not change the results by more than one percent.}
% Conclusions
{}

  
\titlerunning{Asteroseismology of the F-type star 94 Ceti }
  
\authorrunning{Deal et al.}  

\maketitle 

%_____________________________________________________________________
%%%%%%%%%%%%%%%%%%%%%%%%%%%%%%%%%%%%%%%%%%%%%%%%%%%%%%%%%%%%%%%%%%%
\section{Introduction}

Asteroseismic studies of G and F-type stars were strongly boosted by space observations, especially with the \textit{Kepler} satellite. Big data have been obtained, leading to global studies referred to as "ensemble asteroseismology". It is still important however to study some special stars for a deeper insight including detailed internal physical structure. Radial velocity observations of the exoplanet-host star 94 Ceti obtained with HARPS allowed derivation of seismic frequencies precise enough to test the importance of new stellar physics on the derived parameters. This star is on the main sequence, hot enough to suffer in an important way the effects of atomic diffusion including radiative accelerations on heavy elements. This leads to element accumulation in specific internal layers and to extra mixing (Deal et al. 2016). We found interesting to test, at least in a preliminary way, this new physics on an F-type star through asteroseismology. We also wanted to study the impact of introducing a complete atmosphere instead of the usual grey limit.

94 Ceti is a binary system. An exoplanet has been found in orbit around 94 Ceti A, at a distance of 1.42 AU. This star is classified as an F8V star, with a visual magnitude of V = 5.08. Its parallax, $\pi = 44.29 \pm 0.28$ mas (\textit{Hipparcos catalog}, \cite{vanleeuwen07}) puts it at a distance of d = 22.38 pc. Its companion is an M dwarf, which is 6.4 magnitudes fainter at a distance of $\approx$ 100 AU \citep{hale94}.

In the present paper, we first give the results obtained for the parameters of the star 94 Ceti A (HD 19994, HR 962, HIP 14954, GJ 128) by using classical standard evolutionary models and fitting these models with the observations gathered with the spectrograph HARPS. Then we discuss the results obtained with more sophisticated models including detailed radiative accelerations, as computed in the TGEC (Toulouse Geneva Evolution Code). Finally we add complete atmospheres to derive the impact of more sophisticated treatments of the stellar outer layers.

\section{The observational data}

\subsection{Spectroscopic observations}\label{obs}

Several spectroscopic studies have been performed for this star. Some of them suggest that it has a nearly solar metallicity. It is the case of \cite{edvardsson93} who, in their study of the chemical evolution of the galactic disk, gave a metallicity of 0.09. Eight years later, \cite{smith01}, in their work exclusively focused on 94 Ceti A and its abundance distribution, gave a similar metallic value for this star. However most studies, which mainly focused on the exoplanet host-star status of 94 Ceti, lead to a stellar overmetallicity. It is the case for \cite{santos01,santos03,santos04}, who present 94 Ceti among their general studies of statistical properties, metal-rich nature and spectroscopic [Fe/H] for exoplanet-host stars. 

In our study, we use the metallicity values of their 2004 work, which presents a summary and a revision of the previous results, giving separately the values for [Fe/H] derived from the high-resolution spectra obtained with four different instruments: CORALIE, FEROS, UVES and UES. 

\cite{valenti05} also included 94 Ceti in their spectroscopic study of cool stars. Their results are in agreement with Santos et al. results, i.e., this star shows an overmetallicity with respect to the Sun. They also gave estimates of the age, mass and radius of the star. \cite{maldonado12} analysed spectroscopically 94 Ceti as part of their survey about metallicity of solar-like stars with debris disk and planets. Their results are also in agreement with the previous mentioned analyses. 

A summary of the spectroscopic studies previously mentioned is presented in table \ref{table:1}.

\begin{table*}
\centering
\caption{Summary of previous spectroscopic studies of 94 Ceti.} 
\label{table:1}
\begin{tabular}{c c c l}
\hline  
 [Fe/H] & $T_{\mathrm{eff}}$(K) & $\log g$ & References   \\
\hline
$0.09\pm0.10$ & $6104\pm100$ & $4.10\pm0.20$ & \cite{edvardsson93}  \\ 
$0.09\pm0.05$ & $6030\pm20$  & $3.95\pm0.05$ & \cite{smith01}  \\
$0.25\pm0.08$ & $6217\pm67$  & $4.29\pm0.08$ & \cite{santos04} (CORALIE)  \\
$0.32\pm0.07$ & $6290\pm58$  & $4.31\pm0.13$ & \cite{santos04} (FEROS) \\
$0.19\pm0.05$ & $6121\pm33$  & $4.06\pm0.05$ & \cite{santos04} (UVES) \\
$0.21\pm0.08$ & $6132\pm67$  & $4.11\pm0.23$ & \cite{santos04} (UES) \\
$0.19\pm0.03$ & $6188\pm44$  & $4.24\pm0.06$ & \cite{valenti05}  \\
$0.19\pm0.03$ & $6140\pm31$  & $4.35\pm0.09$ & \cite{maldonado12}  \\       
\hline
\end{tabular}
\end{table*}

\subsection{Interferometric observations}

Direct measurements for the radius of 94 Ceti A were done with interferometric techniques by  \cite{vanbelle09}. Using the Palomar Testbed Interferometer (PTI) and the CHARA array, they measured linear radii and effective temperatures of 12 stars with known planetary companions. For 94 Ceti A, a radius of $R/R_\odot =1.898\pm0.070$ was obtained, which gives an important constraint on the stellar modelling. The derived associated effective temperature is $T_{\mathrm{eff}}=6109\pm111$ K.

\subsection{The exoplanet data}

A Jupiter-like companion to 94 Ceti A was first announced by \cite{queloz00}. A more precise orbital solution was later given by \cite{mayor04}. Discovered by using the radial velocity technique, this planet has a minimum mass of 1.68 $M_J$ and orbits its parent star at a distance of 1.42 AU with a period of $\approx$535.7 days. A summary of the properties of HD19994b is given in Table \ref{table:2}.

\begin{table}
\centering
\caption{Summary of HD 19994b properties \citep{mayor04}.}
\label{table:2}
\begin{tabular}{l c }
\hline  
 \textbf{HD 19994b} &    \\
\hline
$M~\sin i$ & $1.68~M_J$   \\ 
Orbital period & $535.7\pm3.1$ days  \\
Semi-major axis & 1.42 AU   \\
Eccentricity & $0.30\pm0.04$  \\
$\omega$ & $211\pm6$  \\      
\hline
\end{tabular} 
\end{table}

\section{Seismic observations and analysis}
\subsection{Mode identification}

The star 94 Ceti A was observed with the HARPS spectrograph mounted on the ESO 3.6 m telescope, at La Silla Observatory, Chile. The observations were carried out during 7 consecutive nights in November 2007. The global Fourier transform of the observations is given in Fig. 1.

The Fourier transform was first analysed with the Period 04 code \citep{lenz05} to extract the most important peaks. However, due to aliasing it may happen that high amplitude peaks are combinations of aliases instead of real modes. For this reason we also investigated manually the Fourier transform, using the window function to pick out the real peaks and eliminate the surrounding aliases. We then used stellar models to identify the modes corresponding to the detected frequencies. This method was quite successful, allowing us to recognize 26 modes in a convincing way.
The frequencies of these identified modes are given in table \ref{table:3} with an uncertainty evaluated as $1\mu$Hz. The average large separation is $\Delta\nu= 64.1 \mu$Hz and the average small separation $\delta\nu_{0,2}= 4.0 \mu$Hz.

\begin{table}
\centering
\caption{Identified P-mode frequencies (in $\mu$Hz) for 94 Ceti.}
\label{table:3} 
\begin{tabular}{c c c c}
\hline  
 $l=0$ & $l=1$ & $l=2$ & $l=3$   \\
\hline
1224 & 1253 & 1218 & 1246   \\ 
1288 & 1316 & 1283 & -   \\
1352 & 1381 & 1347 & -   \\
1415 & 1445 & 1411 & 1438   \\
1479 & 1508 & 1475 & 1501   \\
1545 & 1572 & 1538 & -   \\
1608 & 1637 & -    & 1680   \\
1673 & 1702 & -    & -   \\     
\hline
\end{tabular}
\end{table}

\begin{figure*}
\center
\includegraphics[scale=0.32]{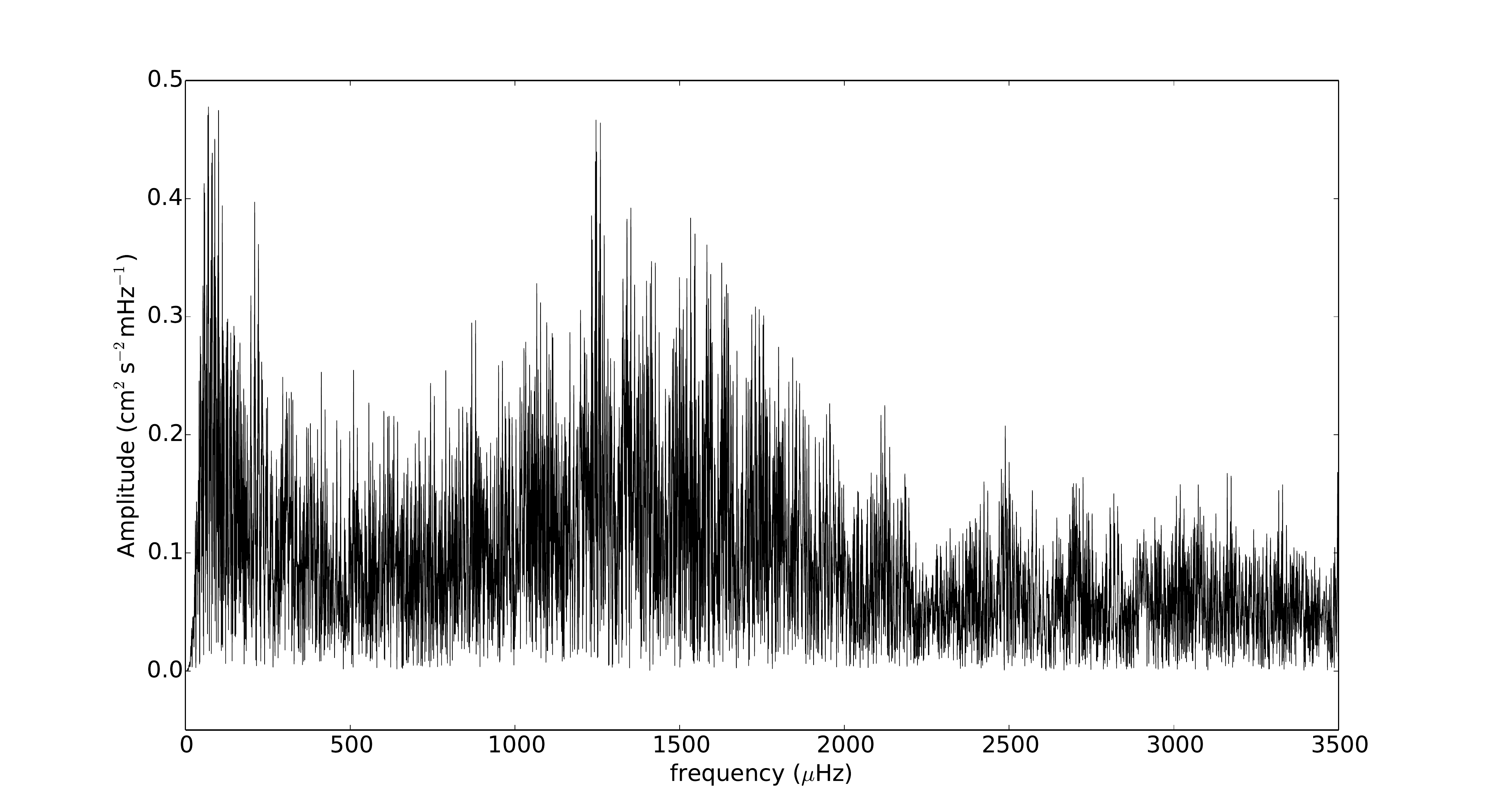}
\label{TF}
\caption{Global Fourier transform of 94 Ceti data.}
\end{figure*}

\subsection{Rotational splitting}

Given the projected rotation velocity $v \sin i$, the rotation period $P_{rot}$, and the linear radius $R_{star}$, the inclination angle $i$ of the rotation axis on the line of sight can be determined. In the case of 94 Ceti, $P_{rot}=12.2$ days (\cite{mayor04}, from $R_{HK}$ activity indicator), $R/R_\odot =1.898\pm0.070$ from interferometric observations \citep{vanbelle09}. We obtain $\sin i \approx 1.02$ which means that the star is seen equator-on. \cite{gizon03} showed that in this case, for $l=1$ modes, the central $m = 0$ component should vanish. With a rotation period of $P_{rot}=12.2$ days, the shift induced by rotational splitting is $\approx 0.95 \mu$Hz. Thus, in the 94 Ceti A case, we expect to see the $m = \pm 1$ components of $l=1$ modes separated by $\approx 1.9 \mu$Hz. For $l=2$ modes, \cite{gizon03} predict the visibility of the three components: $ m = 0$ and $m = \pm 2$. The frequency shift between the visible components should also be $\approx 1.9 \mu$Hz. However, since the predicted amplitudes for the $m = 0$ components are lower than those of the $m = \pm 2$ components, there may be cases where only $m = \pm2$ components are visible. 
The present observations are not precise enough to test these rotational effects, as the uncertainties on the frequency determinations are of order one $\mu$Hz. This may however be the reason for some of the shifts obtained between the observed and computed frequencies as seen in the \'echelle diagram (Fig.3).

Since the star is seen almost equator-on, the mass of the planet HD 19994b should be close to its derived minimum value of 1.68 $M_J$. It may also be transiting in front of the star. To our knowledge, such a transit has not yet been looked for observationally.

\section{Asteroseismic study}
\subsection{Stellar models}

We used the Toulouse Geneva Evolution Code (TGEC) to compute stellar models that fit the seismic observations of 94 Ceti A. This code performs complete computations of atomic diffusion for 21 species, including radiative accelerations for all heavy elements. The 21 species include 12 elements and their main isotopes: H, $^{3}$He, $^{4}$He, $^{6}$Li, $^{7}$Li, $^{9}$Be, $^{10}$B, $^{12}$C, $^{13}$C, $^{14}$N, $^{15}$N, $^{16}$O, $^{17}$O,$^{18}$O, $^{20}$Ne, $^{22}$Ne, $^{24}$Mg, $^{25}$Mg, $^{26}$Mg, $^{40}$Ca and $^{56}$Fe \citep{theado12}. The diffusion coefficients used in the code are those derived by \citet{paquette86}. 

The Rosseland opacities are recalculated inside the model, at each time step and at every mesh point, using OPCD v3.3 and data from \citet{seaton05}, to take into account the local chemical composition. In this way, the stellar structure is consistently computed all along the evolutionary tracks, as well as the individual radiative accelerations of C, N, O, Ne, Mg, Ca, and Fe. This is done by using the improved semi-analytical prescription proposed by \citet{alecian04}. A more detailed discussion of these computations is given in \citet{deal16}.

The equation of state used in the code is the OPAL2001 equation \citep{rogers02}.
The nuclear reaction rates are from the NACRE compilation \citep{angulo99}. The mixing length formalism is used for the convective zones with a mixing length parameter of 1.8, calibrated on the Sun \citep{theado09}.

\subsection{Determination of the best model}\label{bm}

\begin{figure*}
\center
\includegraphics[scale=0.45]{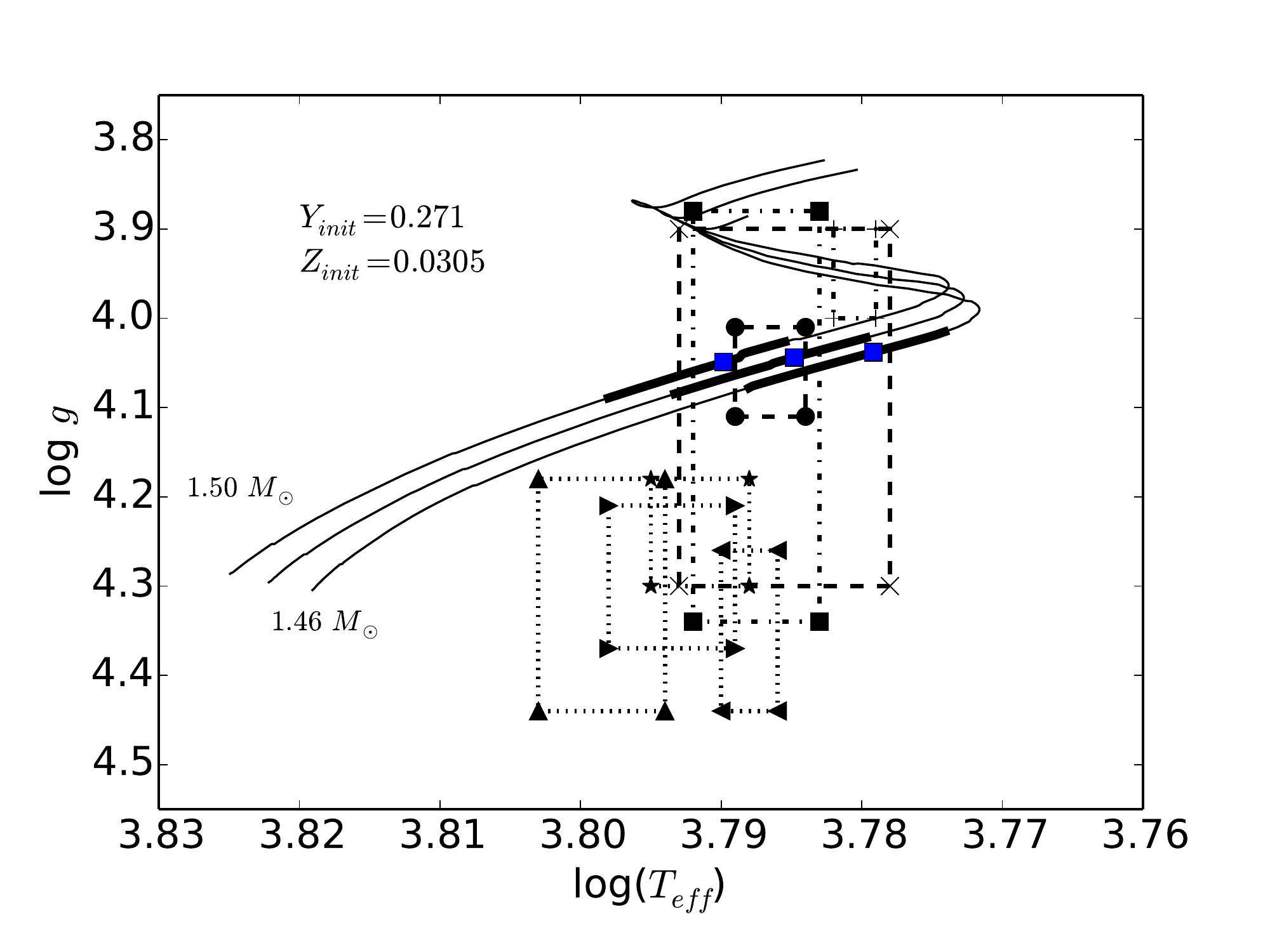}
\includegraphics[scale=0.45]{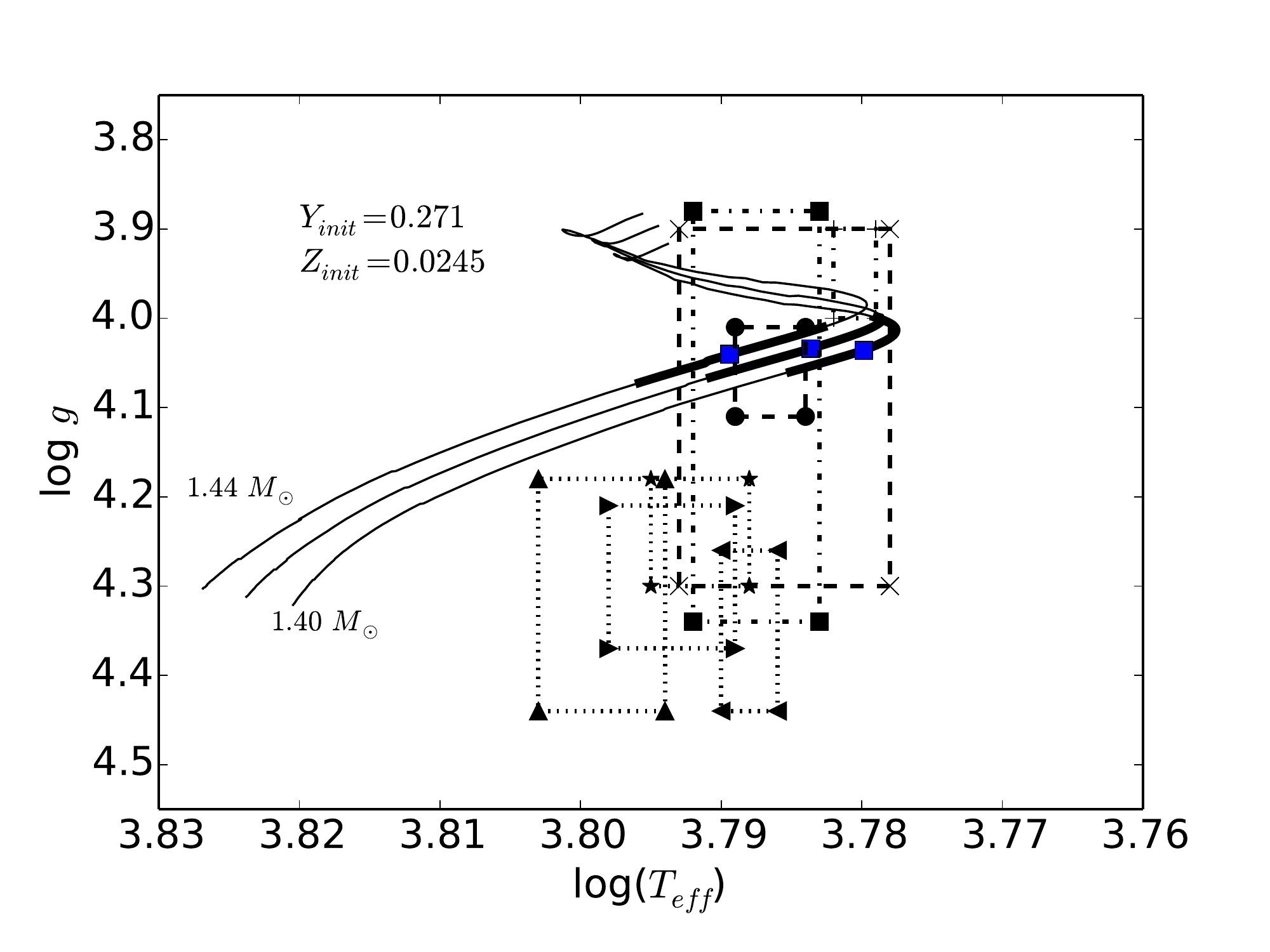}
\includegraphics[scale=0.45]{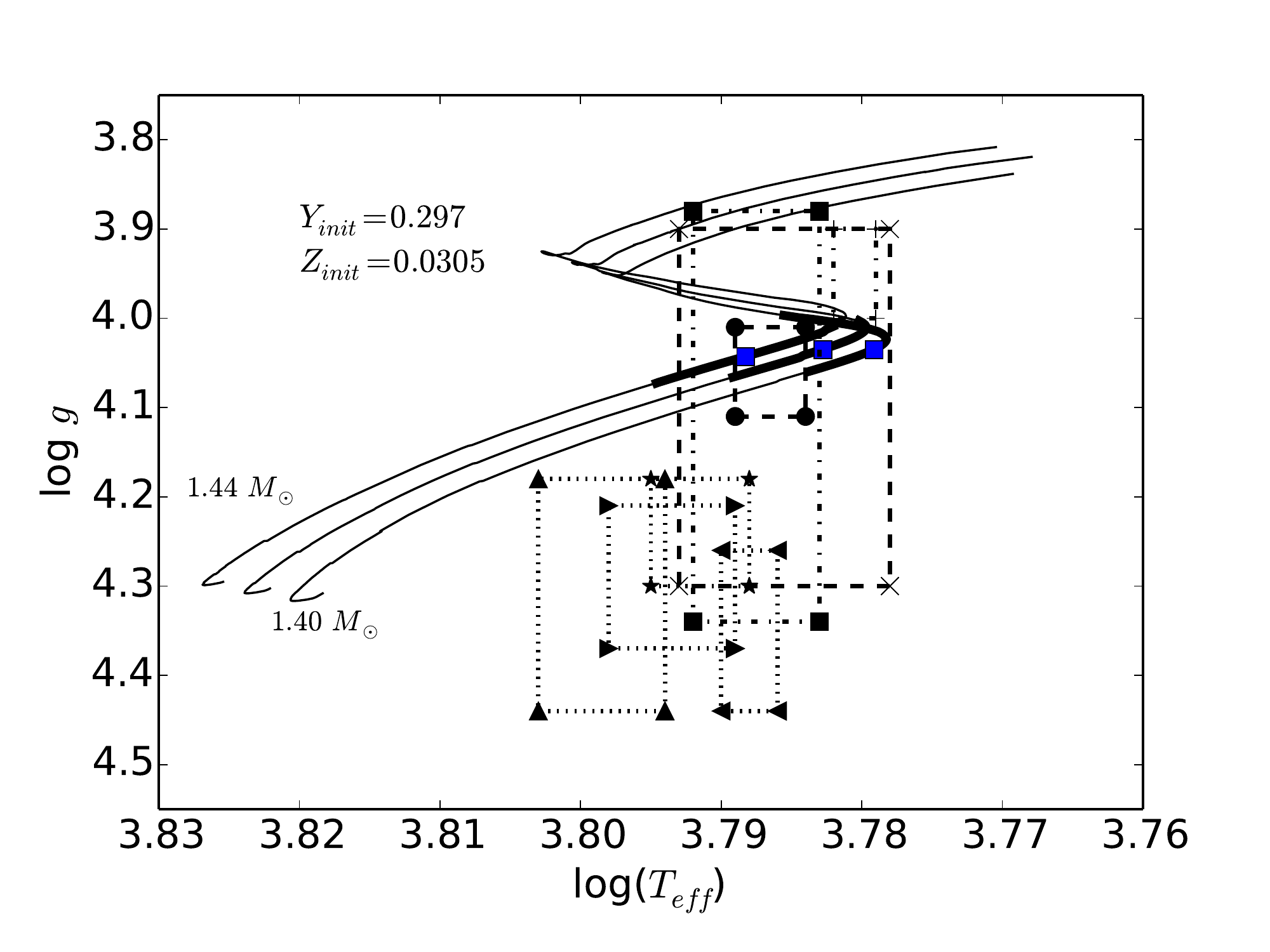}
\includegraphics[scale=0.45]{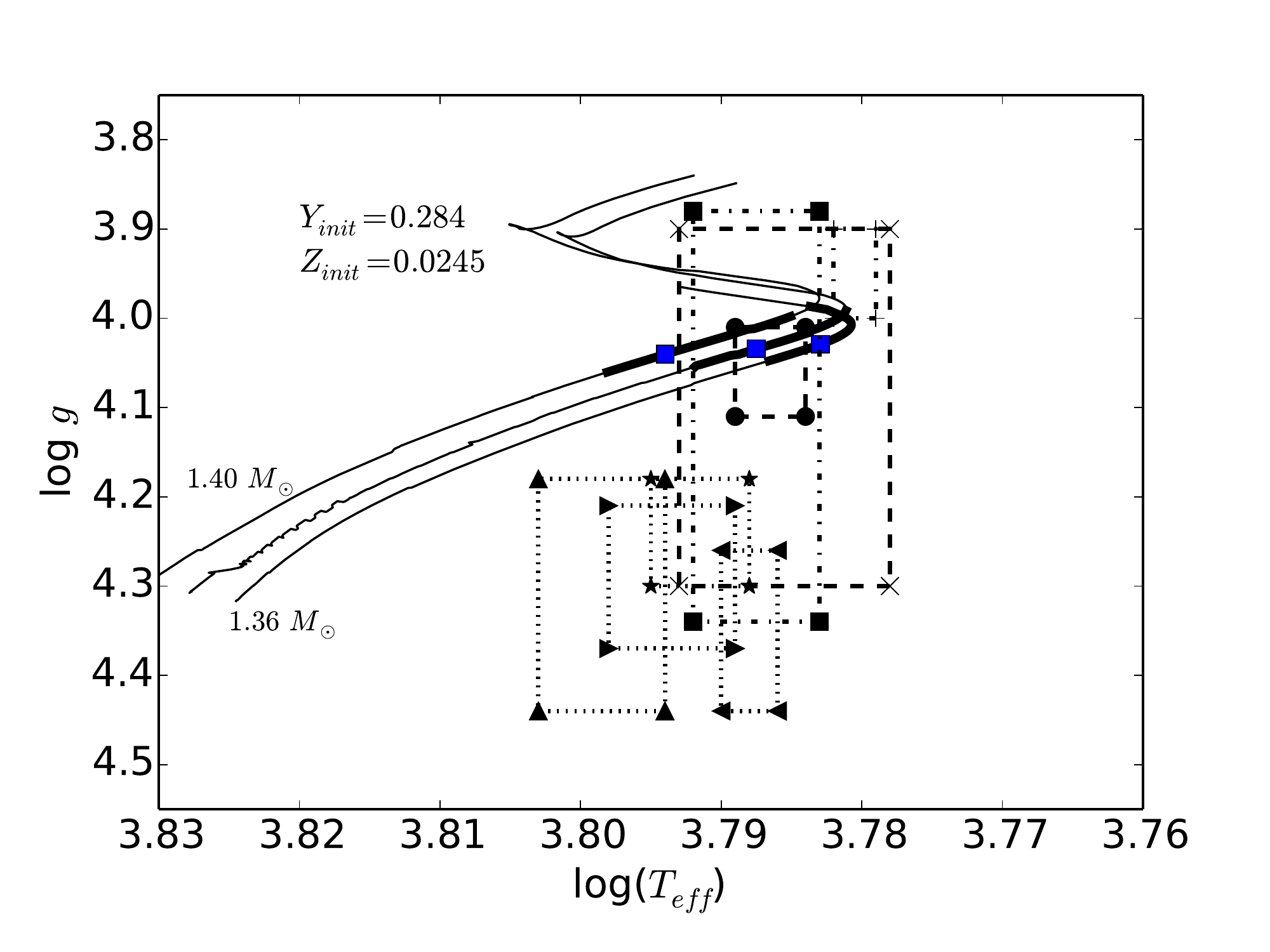}
\caption{$\log g$ vs. $T_{\rm eff}$ diagrams for models with four different initial chemical compositions. The masses range between 1.36 and 1.5~$M_{\odot}$. The error boxes come from the spectroscopic studies detailed in section \ref{obs}, Table 1. The crosses correspond to \cite{edvardsson93}, the plus correspond to \cite{smith01}, the right triangles correspond to \cite{santos04} (CORALIE), the triangles correspond to \cite{santos04} (FEROS), the dots correspond to \cite{santos04} (UVES), the squares correspond to \cite{santos04} (UES), the stars correspond to \cite{valenti05} and the left triangles correspond to \cite{maldonado12}. The blue squares represent the models with the same mean large separation as the one determined from the seismic observations. The heavy line segments correspond to models consistent with the interferometric radius.}
\label{fig2}
\end{figure*}

The stellar oscillation modes were derived for each model using the PULSE code \citep{brassard08}. 
We first computed models including diffusion, without the effects of radiative accelerations, with masses ranging from 1.36 to 1.50 $M_\odot$. The initial helium mass fraction $Y_{init}$ was chosen either as "solar" ($Y_{init}=0.271$, \citealt{grevesse93}) or "galactic" ($Y_{init}=Y_G$), which means that it increases with Z following the chemical evolution of galaxies, as derived by \citealt{izotov04, izotov10}. The initial heavy element mass fraction $Z_{init}$ was chosen so that the surface metallicity at the present age lies inside the error bars of the observed values. This lead to $Z_{init}=0.0245$ for the low metallicity case ([Fe/H]$ = 0.09\pm{0.10}$) and to $Z_{init}=0.0305$  for the high one ([Fe/H]$ = 0.21\pm{0.08}$). The larger metallicities derived by \cite{santos04} with CORALIE and FEROS were not considered. Evolutionary tracks computed for these four different initial compositions are presented in Fig. \ref{fig2}.

We determined for each track the model with the same average large separation as the observed one (represented by blue square) and then we derived the best of all these models for each initial composition. We did this by fitting the \'echelle diagrams with the observed one, using $\chi^2$ minimisations. This selection gave four potential candidates for the best model, one for each initial chemical composition. The first one has a mass of 1.40~$M_\odot$ with $Y_{init}=Y_G$ and $Z_{init}=0.0245$ at 2.32 Gyr (bottom right panel of Fig. \ref{fig2}, model 1), the second one has a mass of 1.44~$M_\odot$ with $Y_{init}=Y_G$ and $Z_{init}=0.0305$ at 2.38 Gyr (bottom left panel of Fig. \ref{fig2}, model 2), the third one has a mass of 1.42~$M_\odot$ with $Y_{init}=0.271$ and $Z_{init}=0.0245$ at 2.79 Gyr (upper right panel of Fig. \ref{fig2}, model 3) and the last one has a mass of 1.44~$M_\odot$ with $Y_{init}=0.271$ and $Z_{init}=0.0305$ at 2.97 Gyr (upper left panel of Fig. \ref{fig2}, model 4). Comparing the $\chi^2$ values, we find that models 1 and 2 best fit the observational values.

An interesting result of this study is the constraint on $\log g$. For all initial compositions used for these model the $\log g$ value ranges between 4 and 4.1. This rules out the spectroscopic $\log g$ and $T_{\rm eff}$ determinations of \cite{santos04} obtained with the instruments CORALIE and FEROS, and \cite{valenti05} and \cite{maldonado12}. On the other hand, the results obtained by \cite{santos04} with UVES and UES are consistent with the seismic determination of $\log g$ .

\begin{figure}
\center
\includegraphics[scale=0.45]{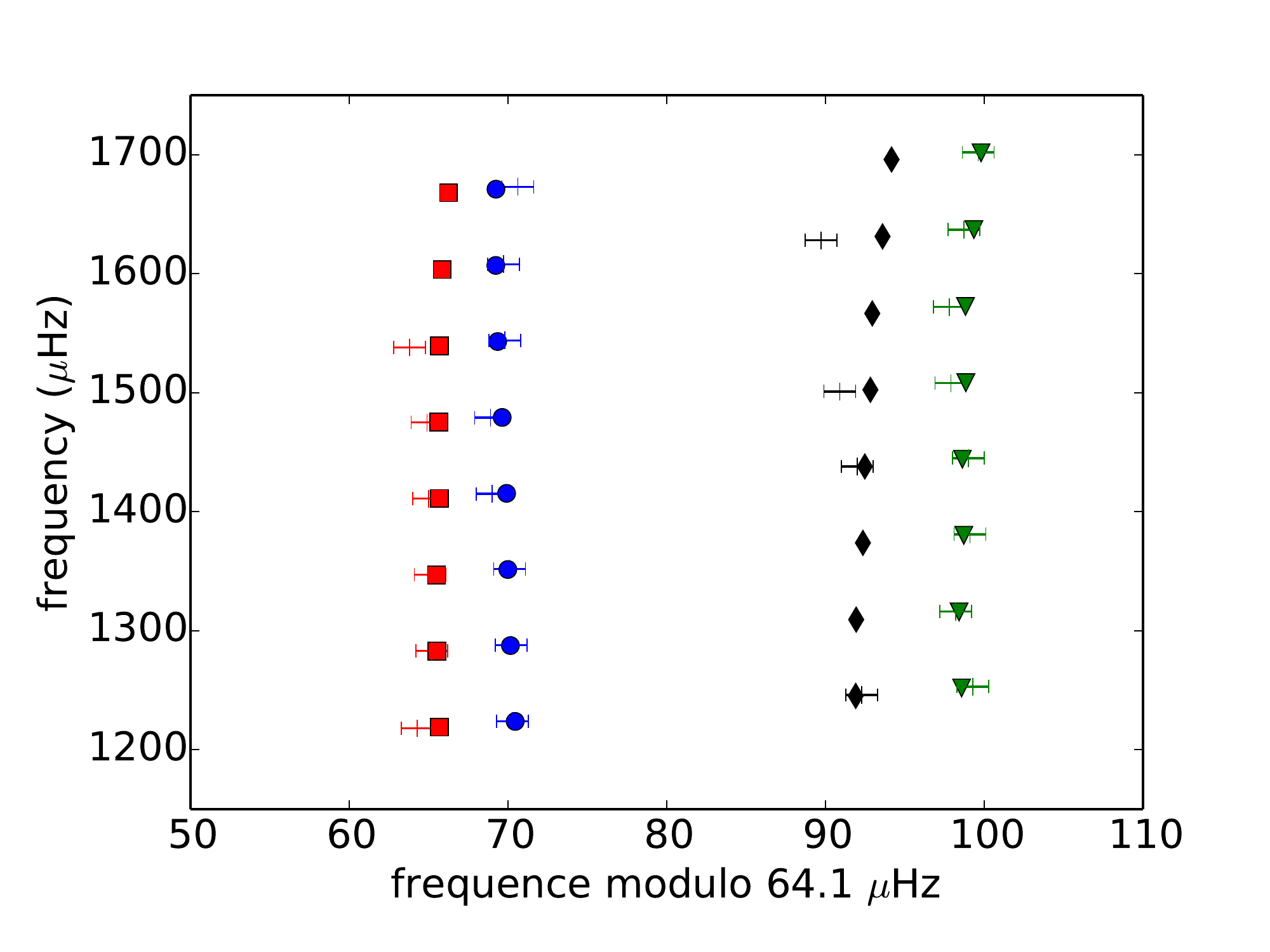}
\caption{Echelle diagram of the 1.44~$M_\odot$ at 2.38 Gyrs. The frequencies computed for the models with $Y_{init}=0.297$ and $Z_{init}=0.0305$ are represented by blue dots (l=0), green triangles (l=1), red squares (l=2), and black diamonds (l=3). The observed frequencies are represented by crosses (error bars : $1\mu$Hz).}
\label{fig3}
\end{figure}

Comparing the seismic results with the spectroscopic observations (see Fig. \ref{fig2}), we find that the best model is model 2, with a mass of 1.44~$M_\odot$, as model 1 lies outside all spectroscopic boxes (see the 1.40~$M_\odot$ model on the bottom right panel of Fig. 2). The \'echelle diagram corresponding to this model is presented in Fig. \ref{fig3}. Its parameters are given in Table \ref{table:4}.

\begin{table}
\caption{Properties of 94 Ceti from this work.} 
\label{table:4}
\centering
\begin{tabular}{c c }
\hline  
 & 94 Ceti    \\
\hline
\smallskip 
$T_{\rm eff}$(K) & $6140$   \\
\smallskip       
log $g$ & $4.041$   \\       
\smallskip
Mass ($M_{\odot}$) & $1.44$   \\        
\smallskip
Radius ($R_{\odot}$) & $1.89$   \\        
\smallskip
Luminosity ($L_{\odot}$) & $0.66$   \\        
\smallskip
Age (Gyr) & $2.38$  \\     
\smallskip
$Z_{init}$ & $0.0305$   \\        
\smallskip
$Y_{init}$ & $0.297$   \\  
\smallskip
$Z_{surf}$\tablefootmark{a} & $0.0268$   \\        
\smallskip
$Y_{surf}$\tablefootmark{a} & $0.243$  \\ 
\smallskip
[Fe/H]\tablefootmark{a} & $0.17$  \\      
\hline
\end{tabular}
\tablefoot{\tablefoottext{a}{Values at the age of best model}
}
\end{table}

\section{Impact of radiative accelerations and fingering convection on the model parameters}

The radiative accelerations are larger than the gravity for stars with masses larger than 1.3~$M_\odot$ at solar metallicity \citep{michaud76}. The best model for 94 Ceti has a mass of 1.44~$M_\odot$. Even though the initial metallicity is larger than solar, the effect of radiative accelerations may be important and has to be investigated. As shown by \cite{richer00} and \cite{richard01}, and more recently by \cite{theado09} and \cite{deal16}, the radiative accelerations may lead to elements accumulation in specific stellar layers, when they are larger than the gravity and decrease towards the surface. These accumulations may also trigger fingering convection if they lead to inversions of $\mu$-gradients \citep{theado09,deal16}. 

Our final aim is to evaluate the differences obtained for the best model parameters when we include or not the radiative accelerations and the fingering convection in the computations. As a first step, we characterise the impact of radiative accelerations and fingering convection on the oscillation frequencies along evolutionary tracks.  
We then chose to compute the difference in ages of our best models, obtained with the same initial conditions of mass and chemical composition.

\subsection{Impact of radiative accelerations }\label{agegrad}

We studied the impact on the age determinations of the star, of including radiative accelerations in the stellar models all along the evolutionary tracks. We used as initial conditions those of the best model derived in Section 4.2. The mean large frequency separations computed in the models were used as a proxy . 

\begin{figure}
\center
\includegraphics[scale=0.45]{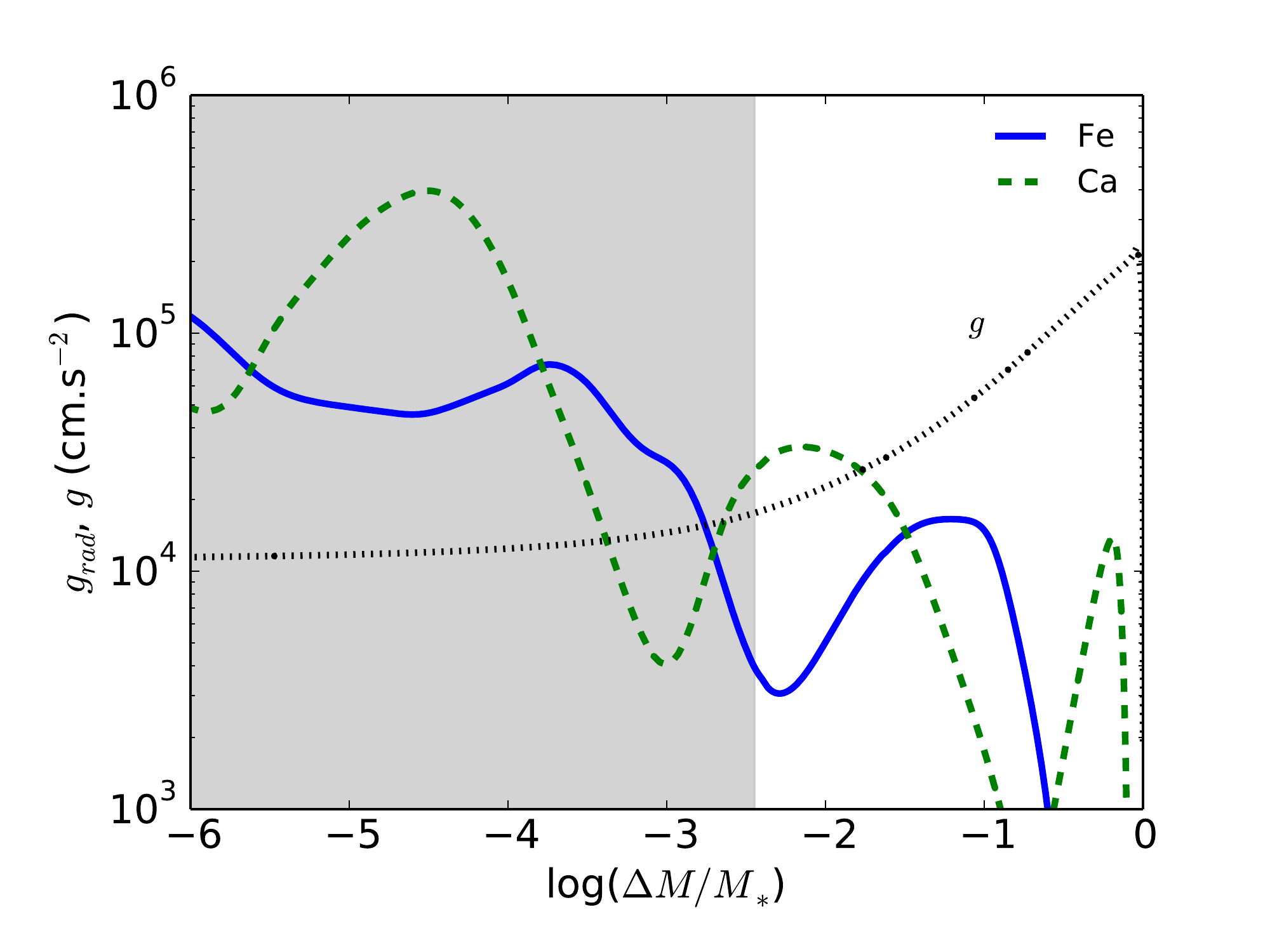}
\caption{Radiative acceleration on calcium (green dashed lines) and iron (blue solid lines) for the model of 1.44~$M_\odot$ at 2.48 Gyr with $Y_{init}=0.297$ and $Z_{init}=0.0305$. The dark grey area represents the surface convective zone.}
\label{fig4}
\end{figure}

\begin{figure}
\center
\includegraphics[scale=0.45]{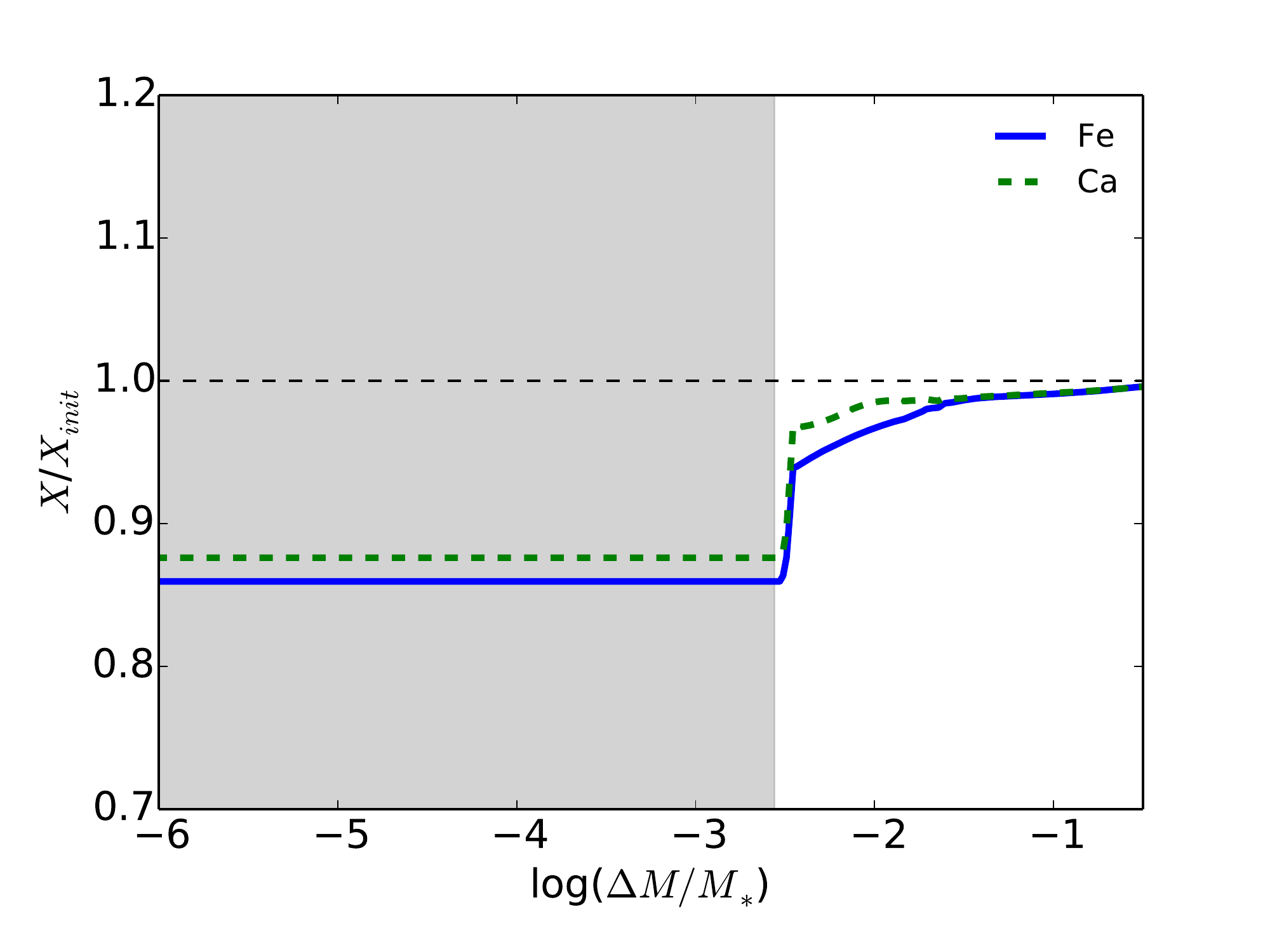}
\includegraphics[scale=0.45]{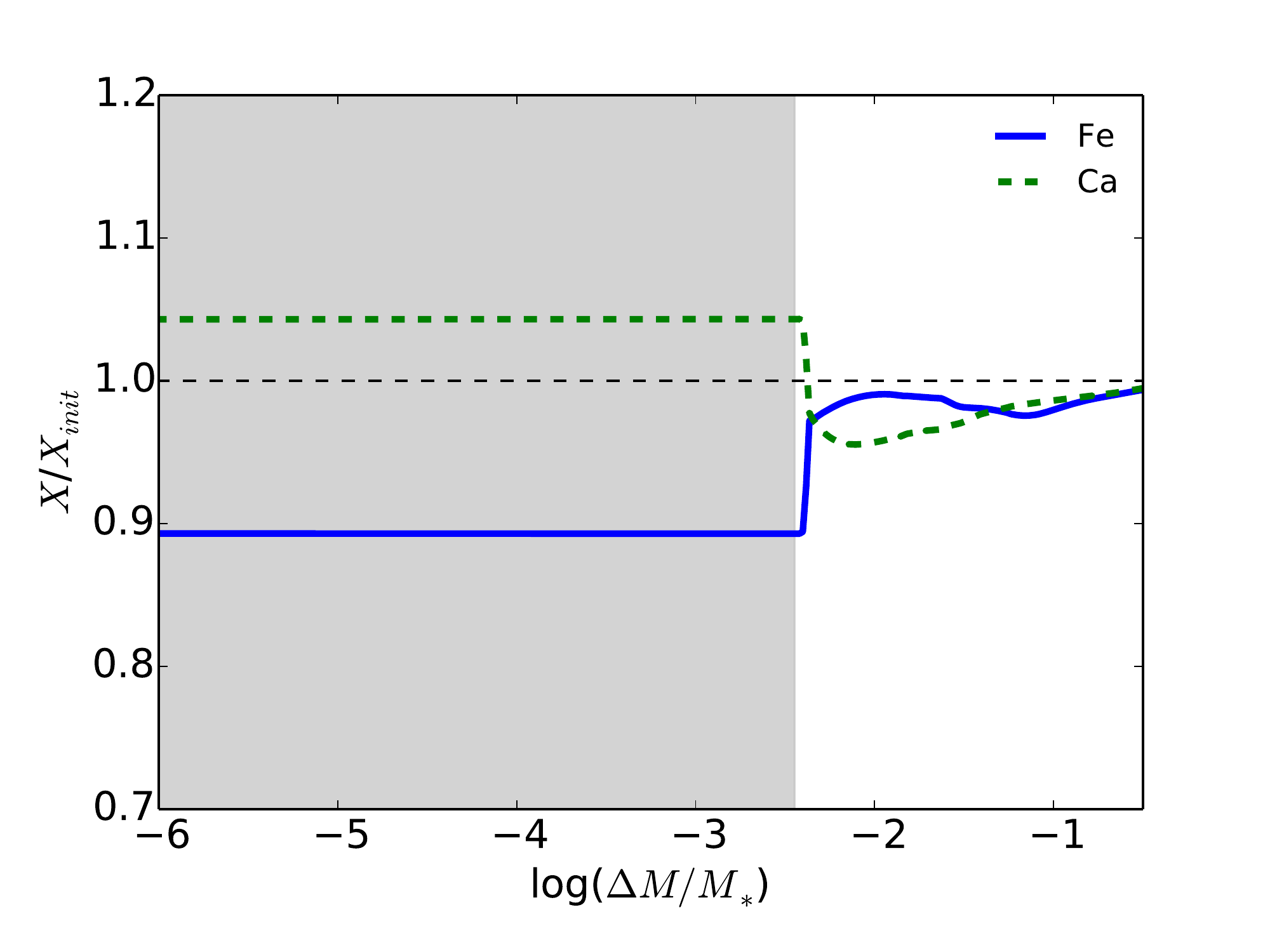}
\caption{Abundance profiles of calcium (green dashed lines) and iron (blue solid lines) for the model of 1.44~$M_\odot$ with $Y_{init}=0.297$ and $Z_{init}=0.0305$ with gravitational settling only (upper panel) and including radiative acceleration (lower panel). The dark grey areas represent the surface convective zone.}
\label{fig5}
\end{figure}

We first computed a 1.44~$M_\odot$ evolutionary track, including radiative accelerations all over the main sequence. The best model that we obtained in this case has an age of 2.48 Gyrs. This is roughly 4\% larger than the age determined without including radiative accelerations. The most important elements in this case are iron and calcium. As we may see on figure \ref{fig4}, the radiative acceleration on calcium is larger than the gravity at the bottom of the surface convective zone. This produces an accumulation of calcium and increases the surface abundance (green dashed curve, lower panel of Fig. \ref{fig5}). 

Furthermore, although the radiative acceleration on iron is lower than the gravity in the radiative zone (solid blue curve of Fig. \ref{fig4}), they are large enough to produce a gradient in the diffusion velocity and to slow down the gravitational settling of iron between $-2<\log (\Delta M/M_\ast)<-1$. This leads to an iron accumulation at the bottom of the surface convective zone (solid blue curve, lower panel of Fig. \ref{fig5}).

\begin{figure}
\center
\includegraphics[scale=0.45]{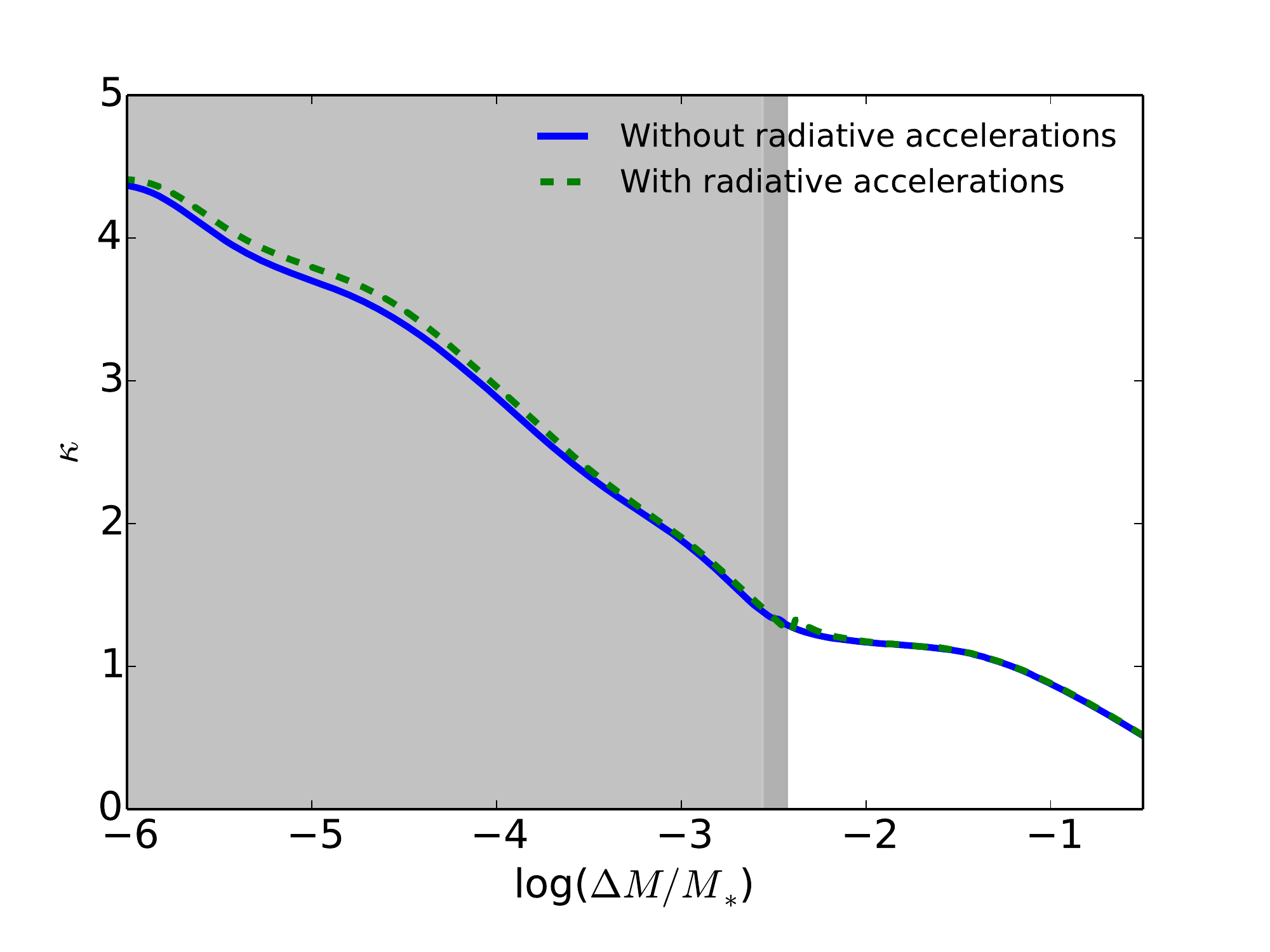}
\caption{Opacity profiles for the model of 1.44~$M_\odot$ with $Y_{init}=0.297$ and $Z_{init}=0.0305$ with gravitational settling only (blue solid curve) and including radiative acceleration (green dashed curve). The light grey area represents the surface convective zone of the model when radiative accelerations are not included. The dark grey area represents the extension of this convective zone obtained when radiative accelerations are included in the computations. }
\label{fig6}
\end{figure}

The abundance profiles are very different when radiative accelerations are taken into account. This modifies the mass of the surface convective zone from $\log (\Delta M/M_\ast)=-2.56$ when taking into account only gravitational settling to $\log (\Delta M/M_\ast)=-2.47$ when including radiative accelerations. This change of the mass of the convective zone by roughly 20\% induces modifications of the oscillation frequencies which partly explains the difference of age obtained between the two models. Another reason is the modification of the opacity profile. Figure \ref{fig6} shows the opacity profiles of the two models. The modification of the abundance profiles changes the opacity in the surface convective zone and at its bottom. This is due to the accumulation of calcium in the surface convective zone and the accumulation of iron around $\log (\Delta M/M_\ast)\approx -2$. These differences in the opacities are only of a few percent, but still have an impact on the mean density of the model and modify the age of the model for which the mean large frequency separation is consistent with the observations.

\subsection{Induced fingering convection}

\cite{theado09} and \cite{deal16} showed that the accumulation of heavy elements in A type stars due to the effect of radiative accelerations may produce an inverse mean molecular weight gradient and trigger fingering (thermohaline) convection. This instability deeply modifies the structure of the star which should have an impact on the oscillation frequencies. In this section we study the impact of the fingering convection on the age determination for 94 Ceti A in the same way as in section \ref{agegrad}.

Fingering convection may occur in stars every time a local inversion of the mean molecular weight gradient appears in the presence of a stable temperature gradient. This may happen for example, in the case of heavy element accumulation or in the case of accretion of metal rich planetary matter \citep{vauclair04,garaud11,deal13,deal15}.

Fingering convection is characterised by the so-called density ratio $R_0$ , which is the ratio between the thermal and $\mu$-gradients: 
\[
R_0=\frac{\nabla - \nabla_{ad}}{\nabla_{\mu}}. 
\] 
The instability can only develop if this ratio is higher than one and lower than the Lewis number, which is the ratio of the thermal to the molecular diffusivities. In this case, a heavy blob of fluid falls down inside the star and continues to fall because it exchanges heat more quickly than particles with the surroundings. If $R_0$ is smaller than one, the region is dynamically convective (Ledoux criterium), and if it is larger than the Lewis number, the region is stable.

Various analytical treatments of fingering convection in stars were given in the past, leading to mixing coefficients that differed by orders of magnitude \citep{ulrich72,kippenhahn80}. More recently, 2D and 3D numerical simulations were performed, converging on coefficients of similar orders \citep{denissenkov10,traxler11}. 

Here we tested the occurence of fingering convection by using the recent prescription given by \citet{brown13}, which has been confirmed by the 3D simulations of \cite{zemskova14}. Whereas fingering convection may occur later on the evolutionary track, we found that in our best model for 94 Ceti the radiative layers are not yet subject to this instability.

The F-type star 94 Ceti lies in the range of effective temperatures for which radiative accelerations on individual elements begin to have important effects on the stellar internal structure. In a general way, the effects are larger for larger effective temperatures and lower metallicities. Many stars observed in the future by space missions like TESS or PLATO will be affected by such effects. The characterization of exoplanets needs very precise stellar parameters as the mass, the radii and the age. With the precision of these new instruments, it will be necessary to always perform complete computations of atomic diffusion including radiative accelerations as well as the induced mixing processes, for comparison with the observations. 

\section{Effect of modifying the atmospheric boundary conditions}

When computing acoustic frequencies in stellar models obtained with the usual evolution codes, the waves are assumed completely reflected at some layer in the atmosphere, which is a crude approximation, becoming worse and worse for larger frequencies, close to the cut-off value. This has been widely discussed in the literature, and a recipe has been given by \cite{kjeldsen08} to try correcting that effect. 

For the present paper, we choose to compute one complete model including a realistic atmosphere and compare the results with those of the same model computed with the usual atmospheric approximations. This is not trivial and we discuss below how we proceed.

In stellar interiors, the medium is optically thick, and the radiative transfer is computed using the diffusion approximation. The outermost layers of a star being an optically thin medium, their structure implies the detailed computation of the radiative transfer. The approach we choose is to merge 
the internal structure of a TGEC model with a detailed ATLAS9 \citep{kurucz93} model atmosphere, in order to show the difference obtained on the seismic properties.

\subsection{Computation of the complete model}\label{calc_atm}

We present here the main steps we use to perform the junction between a stellar interior and a realistic stellar atmosphere (see \cite{hui-bon-hoa13} for more details).

For a given TGEC model, we compute a model atmosphere with the same surface chemical composition and fundamental parameters. We merge the atmosphere and the interior at the layer where the temperature at the bottom of the atmosphere matches that of the interior. Once the atmosphere is anchored to the internal structure, slight adjustments are made to ensure continuity for the pressure. The density is then recomputed accordingly, assuming a perfect gas law. In the overlapping regions, the transition for each physical quantity from the interior to the atmospheric value is done by using a third degree polynomial weighting.

The mixing length parameter $\alpha_{MLT}$ is usually slightly larger in the internal structure than in the atmosphere, where observational constraints can lead to values as low as 0.5 \citep{veer-Menneret96}. In our case, we have checked that modifying the atmospheric mixing length parameter leads to negligible effects on the computed frequencies.

\subsection{Effect on the mean large frequency separation}

%\begin{figure}
%\center
%\includegraphics[scale=0.45]{ED_atm.eps}
%\caption{Echelle diagrams of the 1.44~$M_\odot$ at 2.37 Gyrs with a full atmosphere. The frequencies computed for the models with $Y=0.297$ and $Z=0.0305$ are represented by blue dots (l=0), green triangles (l=1), red squares (l=2), and black diamonds (l=3). The observed frequencies are represented by crosses (error bars : $1\mu$Hz).}
%\label{fig9}
%\end{figure}

We built a complete stellar model in this way, using the best TGEC model found in Sect.~5.2. This model has a mass of 1.44 $M_{\odot}$, an initial composition $Y_{init}=0.297$ and $Z_{init}=0.0305$, an effective temperature $T_{\rm eff} = 6142~K$ and a gravity $\log g$ =4.0. The average large separation found for the model without complete atmosphere was $64.1~\mu Hz$. Computing the frequencies and separations in the new model including complete atmosphere leads to a slightly smaller average large separation, namely $63.8~\mu$Hz. This difference, smaller than $1~\mu$Hz, is mainly due to the more realistic location of the outer turning point. Instead of being artificially set at the outermost layer of the internal structure model, it now occurs at the place where the atmosphere becomes radiative, that is, where the Brunt-V\"ais\"al\"a frequency rises again. The \'echelle diagram obtained in this case is very similar to the one drawn from the original model.
We thus conclude that the results obtained in previous sections are robust.

\section{Conclusion}

The observations of 94 Ceti A made with HARPS allowed to obtain precise parameters for this star. Several error boxes as given by spectroscopists have been excluded on this basis. Our best model lies inside the boxes given by \cite{santos04}, obtained with UVES and UES observations. On the other hand, those obtained with FEROS and CORALIE give a gravity which is too large and an effective temperature too high. 

We have tested the effect of including atomic diffusion with radiative accelerations in the models, as well as the induced fingering convection, on the final results. The star 94 Ceti A is hot enough to have important radiative effects, in spite of its high metallicity which reduces them. We found that the age of the final best model is a few percents larger than obtained without introducing radiative accelerations in the computations. This difference should be larger for hotter stars and for stars with lower metallicities.

Note that this age difference may still be larger than the one derived in the present paper. We computed radiative accelerations by using SVP tables \citep{alecian04} in which some elements are not yet included. Nickel, for example, may be important in that respect (see \citealt{turcotte98}). 

Introducing a complete ATLAS atmosphere for the computation of the best model does not change the results in a significant way (less than one percent).

Considering the precision obtained with present and future space missions, our results show that complete atomic diffusion should be introduced in the models for stars with masses larger than 1.4 solar masses.

\begin{acknowledgements}
We thank the referee for his/her important remarks on the first version of this paper.
The observations were done with the HARPS spectrograph, ESO program number 080.D-0408. We also thank the French Programme National de Physique Stellaire (PNPS) of INSU/CNRS for financial support.
This work was performed using HPC ressources from GENCI-CINES/IDRIS and from LUPM. 
\end{acknowledgements}  
  
%%%%%%%%%%%%%%%%%%%%%%%%%%%%%%%%%%%%%%%%%%%%%%%%%%%%%%%%%%%%%%%%%%%  
\bibliographystyle{aa} % style aa.bst
\bibliography{biblio.bib} % your references Yourfile.bib  
%%%%%%%%%%%%%%%%%%%%%%%%%%%%%%%%%%%%%%%%%%%%%%%%%%%%%%%%%%%%%%%%%%%
%_____________________________________________________________________

\end{document}